\documentclass{article}

\usepackage{arxiv}
\usepackage{amsmath}
\usepackage[utf8]{inputenc} % allow utf-8 input
\usepackage[T1]{fontenc}    % use 8-bit T1 fonts
\usepackage{caption}
\usepackage{float}
\usepackage{hyperref}       % hyperlinks
\usepackage{url}            % simple URL typesetting
\usepackage{booktabs}       % professional-quality tables
\usepackage{amsfonts}       % blackboard math symbols
\usepackage{nicefrac}       % compact symbols for 1/2, etc.
\usepackage{microtype}  
\usepackage{multirow}% microtypography
\usepackage{array}
\usepackage{xcolor}
\usepackage{bbm}
\usepackage{lipsum}
\usepackage{comment}
\usepackage{graphicx}
\graphicspath{ {./images/} }
\DeclareMathOperator*{\argmax}{arg\,max}

\title{A Review of Statistical Methods for Spontaneous Reporting System Data Mining: Signal Detection and Beyond}

\date{ }

\author{
 Yihao Tan \\
  Department of Biostatistics\\
  School of Public Health and Health Professions\\
  State University of New York at Buffalo\\
  Buffalo, New York, USA\\
   \And
 Marianthi Markatou\thanks{Joint senior and corresponding authors} \\
  Department of Biostatistics\\
  School of Public Health and Health Professions\\
  State University of New York at Buffalo\\
  Buffalo, New York, USA\\
  \texttt{markatou@buffalo.edu} \\
  \And
 Saptarshi Chakraborty$^*$ \\
  Department of Biostatistics\\
  School of Public Health and Health Professions\\
  State University of New York at Buffalo\\
  Buffalo, New York, USA\\
  \texttt{chakrab2@buffalo.edu} \\
}

\begin{document}
\maketitle
\begin{abstract}
Postmarketing safety surveillance relies on data from spontaneous reporting systems (SRS) such as FAERS, EudraVigilance and VigiBase, and commonly uses SRS data mining methods to assess the associations between drugs and adverse events (AEs). Traditionally, these analyses have focused on signal detection framed as a binary decision problem, whereas more recent work has emphasized more nuanced inference involving signal strength estimation and  uncertainty quantification. In this paper, we review contemporary SRS data mining approaches and their statistical underpinnings for safety assessment using data from major pharmacovigilance databases worldwide. In addition to methodological review, we provide practical guidance on data preprocessing for such analysis, including construction of SRS contingency tables using only aggregated AE-drug counts, as are publicly available from databases such as VigiBase and EudraVigilance. We illustrate the guidance via opioid-related datasets obtained from FAERS and VigiBase, complied with subsequent downstream SRS data analyses.
\end{abstract}

% keywords can be removed
%\keywords{First keyword \and Second keyword \and More}

\section{Introduction}

Medicines are essential for disease prevention and treatment; yet, they may also cause undesirable or unexpected adverse events (AEs). Pharmacovigilance is the science dedicated to detecting, assessing, understanding, and preventing adverse effects and other complications associated with medicines, vaccines, medical devices, therapeutic biologics, and other medical products. It plays a crucial role in addressing these risks \cite{WHOPharma}. Postmarket medical product safety assessment, a central component of contemporary pharmacovigilance, focuses on detecting and analyzing AEs once products enter the market. Although clinical trials - the gold standard for medical product safety assessment - identify many AEs, they are often limited by size, duration, scope, and ethical considerations, and may not capture the variability and complexity of the real-world patient populations. Consequently, the analysis of postmarketing surveillance data provides an essential and complementary approach to clinical trials for identifying, characterizing, and monitoring these events. We focus on the discussion on adverse event detection and inference in the context of medicines (i.e., drugs).

Postmarketing surveillance primarily relies on the collection of spontaneous reports of suspected adverse drug events from pharmaceutical companies, healthcare professionals, and patients. These reports are curated and stored in spontaneous reporting system (SRS) databases, such as the US FDA’s Adverse Event Reporting System (FAERS) \cite{FDA_FAERS}, the European Medicines Agency’s (EMA) EudraVigilance \cite{EudraVigi}, and the World Health Organization’s (WHO) VigiBase \cite{VigiBase}. Because SRSs depend on the passive receipt of reports, they are inherently observational, which leads to several statistical challenges in their data analysis. These include underreporting of AEs, absence of appropriate controls, limited information on drug exposure, and susceptibility to selection bias and confounding \cite{markatou2014pattern}. These limitations preclude higher-level insights, such as causal inference from being drawn from SRS data, and restrict the scope primarily to AE-drug association learning. Although some alternatives to SRSs have been proposed to address some of these challenges – e.g., the FDA Sentinel Initiative in the U.S. implements a surveillance framework based on health insurance claims data for signal identification \cite{davis2023use} - SRS databases remain a major source of real-world evidence on drug-related AEs, and have been shown to provide important safety signals when appropriately analyzed.

In this review paper, we focus on SRS data and associated, commonly used statistical methods that have been developed for AE safety signal assessment through AE-drug association analysis. A common underlying principle of these methods is to compare the observed frequency ($O$) of a specific AE-drug pair against its expected null baseline frequency ($E$), where $E$ represents the theoretical expected count that would arise in the absence of any association between the AE and the drug. Estimation of this expected count requires the marginal total number of reports involving the drug and the marginal total number of reports involving the AE. Unlike a traditional epidemiological case-control study, however, SRS databases catalog data exclusively from drug users and do not include natural controls (individuals who did not use any drug); this precludes exact estimation of the marginal AE total counts. Instead, the SRS data mining methods usually approximate this total number of AE occurrences using the best proxy available within an SRS database, namely by pooling information across a broad range (possibly all) of drugs cataloged in the database – including those unrelated to the primary drug(s) of interest – as  enumerated either across the entire database or within defined subsets, such as all reports submitted between 2020 and 2025.

Currently, the central focus of SRS data mining is \textit{signal detection}, which seeks to determine whether the observed frequency ($O$) for an AE-drug pair substantially exceeds the corresponding null baseline count $E$, or equivalently, whether its signal strength, defined as the $O/E$ ratio or a related parametrization, is significantly greater than 1. AE-drug pairs meeting this criterion are considered signals; all others are deemed non-signals. However, contemporary biomedicine and science as a whole have increasingly recognized the limitations of this simplistic signal/non-signal dichotomy, which can obscure important data nuances. While some of the statistical methodologies proposed in pharmacovigilance over the past few decades are capable of providing more nuanced inference beyond binary signal detection – particularly when appropriately adjusted to balance flexibility and regularized estimation – explicit emphasis on uncertainty-aware signal strength estimation has emerged only recently. For example, Tan et al. \cite{tan2025flex} introduced a nonparametric empirical Bayes framework to SRS data mining that provides a data-adaptive estimated posterior distribution of the signal strength parameter ($O/E$) for each AE-drug combination, enabling both flexible estimation and formal uncertainty quantification. This approach supports more nuanced statistical inference for AE-drug associations in several ways. First, signal strength estimation: one can rigorously differentiate between AE-drug combinations with similar signal detection outcomes, such as those with estimated signal strengths of 1.5 and 4.0, which would otherwise both be classified simply as signals within a dichotomized framework. Second, uncertainty quantification: even if two drug-AE pairs have similar expected signal strengths of 1.5, their posterior distributions may differ substantially; for example, 95\% equi-tailed credible intervals of (1.05, 4.95) versus (1.45, 1.65) indicate markedly greater reliability of the estimated signal strength for the second pair. Such distinctions are not accessible under traditional binary signal detection approaches.

However, statistical analysis for SRS data presents several methodological statistical challenges. 1) High dimensionality: An SRS database typically records occurrences of many AEs across a very large number of drugs. The resulting AE-drug contingency table can be extremely high-dimensional, involving thousands of drugs and AEs. This high dimensionality not only makes data modeling computationally intensive but also introduces issues of signal sparsity. 2) Zero-inflation: A notable feature of SRS data is the overwhelming presence of zero counts. Many zeros are structural zeros arising out of AE-drug pairs that are biologically or physically impossible to co-occur (e.g., an oral medication and an injection-site reaction). Other zeros represent unreported but possible combinations, termed sampling zeros. The commonly used ordinary Poisson model, which assumes that zeros arise solely by chance, cannot adequately capture this dual mechanism. 3) Heterogeneity of signal strength: The signal strength for each AE-drug pair varies considerably across SRS data.

In recent decades, a variety of methods have been proposed for SRS data mining, with increasing model complexity over time, and increasingly more complex models being proposed in recent times to adequately handle the complexity and capture nuances of SRS data. Early stage methods, such as proportional reporting ratios (PRR)\cite{evans2001use} and reporting odds ratio (ROR)\cite{rothman2004reporting}, due to limited computational resources, typically modeled each AE-drug association of interest based on corresponding collapsed $2 \times 2$ contingency tables constructed from the SRS data. These methods, commonly referred to as disproportionality analysis. These methods consider only the counts for a drug of interest and all other drugs combined along with an AE of interest and other AEs combined are considered (see Table \ref{table:collapsed-contin-table} in Section \ref{sec:method_review} where a detailed description of construction of SRS table is provided). In this setting, each AE-drug pair is analyzed separately together with its marginal information (i.e., `other AEs' and `other drugs'), and signals are identified by applying ad-hoc thresholds directly to the resulting statistics. 

%Evans et al.  \cite{evans2001use} proposed to use proportional reporting ratios (PRR) for collapsed $2 \times 2$ contingency tables to detect signals for SRS data. Rothman et al.  \cite{rothman2004reporting} pointed out the limitations of PRR and suggested using the reporting odds ratio (ROR) instead. Currently, ROR is the signal detection method recommended by the European Medicines Agency (EMA) used for EudraVigilance\cite{EMA_ROR}.  The Bayesian confidence propagation neural network (BCPNN) \cite{bate1998bayesian}, although based on a parametric Bayesian model for the entire SRS table, produces posterior inferences (Information Component) that depend only on the corresponding AE, drug, and their marginals; hence, the BCPNN can also be regarded as a method based on collapsed $2 \times 2$ table. 

The Poisson model-base original likelihood ratio test (LRT) method, introduced by Huang et al. \cite{huang2011likelihood}, provided a formal frequentist framework for signal detection by comparing observed counts and expected null baseline counts under a Poisson model, which ensures optimal data-driven cutoffs with controlled type-I error and false discovery rate (FDR) while maintaining sufficient power and sensitivity. Subsequent work by Ding et al. \cite{ding2020evaluation} adapted this framework to distinguish between AE-based and drug-based formulations, demonstrating superior performance over traditional signal detection metrics. Recognizing the limitations of the Poisson assumption in sparse datasets with excess zeros, Huang et al.\cite{huang2017zero} adopted a zero-inflated Poisson (ZIP) model with the LRT methodology, yielding the zero-inflated Poisson (ZIP) based LRT (ZIP-LRT). Furthermore, Zhao et al.\cite{zhao2018extended} developed extended versions of both the LRT and ZIP-LRT to enable inference across drug classes while preserving type-I error control. More recently, a relative reporting rates-based parametrization has been proposed, leading to pseudo-LRT frameworks\cite{chakraborty2022use} that have been shown to improve computational efficiency and power in zero-inflated settings.

On the Bayesian side, Bate proposed the first Bayesian method for signal detection, the Bayesian Confidence Propagation Neural Network (BCPNN) \cite{bate1998bayesian}, which utilizes an Information Component ($IC$) to identify AE-drug signals. The Gamma-Poisson Shrinker (GPS)\cite{dumouchel1999bayesian} and multi-item Gamma-Poisson Shrinker (MGPS)\cite{dumouchel2001empirical} methods brought empirical Bayes modeling into pharmacovigilance by deploying a two-gamma mixture prior under a Poisson model, which jointly models the signal/non-signal mixture nature of signal strength of AE-drug pairs for the entire SRS table. Heo and Jung \cite{heo2020extended}  extended GPS to account for structural zeros. Hu et al. \cite{hu2015signal} further generalized the GPS model by leveraging a hierarchical Dirichlet process (HDP) mixture of Poisson distributions, resulting in an infinite mixture prior that flexibly accommodates heterogeneity of signal strength. Tan et al.\cite{tan2025flex} introduced a comprehensive nonparametric empirical Bayes framework, adapting the Koenker and Mizera (KM) nonparametric empirical Bayes method\cite{koenker2014convex}, Efron’s nonparametric empirical Bayes approach\cite{efron2016empirical}, and a general-gamma-mixture method for use in pharmacovigilance. This framework offers flexible empirical Bayes modeling that can effectively handle zero-inflation and heterogeneity of signal strengths, while remaining computationally scalable, and providing superior signal detection and signal strength estimation as evidenced in extensive simulations. Other approaches in SRS data mining include the pattern discovery method MDDC \cite{liu2025mddc}, sequential testing-based methods \cite{li2021vaccine, shih2010sequential}, and an adaptive LASSO-based approach \cite{courtois2021new}. We summarize the freely available SRS data mining software packages in R and Python in Table \ref{table:method_accessibility}.

\begin{table}[ht]
\centering
\caption{Accessibility of methods in SRS data mining}\label{table:method_accessibility}
\begin{tabular}{llll}
\hline
Package name  & Implemented methods & Package repository & Aim \\ \hline
PhViD \cite{ahmed_phvid_2016}    & PRR, ROR, BCPNN,GPS      & CRAN &    Signal detection        \\
openEBGM \cite{pkg_openEBGM, canida2017openebgm} & GPS & CRAN   &    Signal detection    \\
pvLRT \cite{pkg_pvlrt, pvlrt-packagepaper} & (pseudo) LRT approaches  & CRAN & Signal detection \\

 &based on log-linear models &&\\
 sglr \cite{pkg_sglr} &Sequential Generalized   &CRAN &Signal detection \\
 &Likelihood Ratio && \\
 &decision boundaries &&\\
 Sequential \cite{pkg_Sequential} &  Max SPRT statistic & CRAN & Signal detection \\
 AEenrich \cite{pkg_AEenrich} &Modified Fisher’s exact test  & CRAN &Signal detection\\
 & and Modified Kolmogorov  &&\\ 
 &Smirnov statistic && \\
 MDDC \cite{pkg_MDDC, liu2025mddc} &Modified Detecting Deviating & CRAN, PyPI & Signal detection\\
 &Cells Algorithm &&\\
 adapt4pv \cite{pkg_adapt4pv} &adaptive lasso based & CRAN & Signal detection\\
 &signal detection methods &&\\
 pvEBayes \cite{pkg_pvEBayes, tan2025pvebayes} & GPS, K-gamma, General-gamma,  &CRAN & Signal detection and\\
 &KM and Efron && signal strength estimation\\
\hline
\end{tabular}
\end{table}

The remainder of the article is organized as follows. Section 2 reviews the statistical approaches used in pharmacovigilance. Section 3 provides a comprehensive discussion of the three largest pharmacovigilance databases: FAERS, EudraVigilance, and VigiBase. In this section, we propose a procedure for constructing SRS datasets using publicly accessible data from EudraVigilance and VigiBase, and we demonstrate the application of SRS data mining methods to the resulting datasets. Finally, Section 4 concludes the article with a discussion.

\section{Statistical methods in pharmacovigilance}
\label{sec:method_review}

This section briefly reviews major SRS data mining methods proposed in the literature, including PRR, ROR, LRT-based methods, the Gamma-Poisson Shrinker (GPS), the hierarchical Dirichlet process (HDP) model, and nonparametric empirical Bayes approaches. The review is not intended to be exhaustive but aims to summarize key methodological advances in this field over the past decades. We consider an SRS dataset cataloging AE reports on $I$ AE rows across $J$ drug columns. As shown in Table \ref{table:SRS-frequency-table}, let $N_{ij}$ denote the number of reported cases for the $i$-th AE and the $j$-th drug, where $i = 1,..., I$ and $j = 1,..., J$. Therefore, AE-drug pairwise occurrences from the AE-reports are summarized into an $I \times J$ contingency table, where the $(i,j)
$-th cell catalogs the observed count $N_{ij}$ indicating the number of cases involving the $i$-th AE and the $j$-th drug. Some early methods model the AE-drug association by collapsing the SRS table into a $2 \times 2$ contingency table for each AE-drug occurrence ($N_{ij}$) along with the related marginal counts $N_{i\bullet} = \sum_{j=1}^J N_{ij}$, $N_{\bullet j} = \sum_{i=1}^I N_{ij}$ and $N_{\bullet \bullet} = \sum_{i=1}^I \sum_{j=1}^J N_{ij}$ as showed in Table \ref{table:collapsed-contin-table}. Since an SRS table contains nonnegative integer entries in each cell, the Poisson model is a natural choice for modeling the cell counts in SRS data mining methods. An AE-drug occurrence $N_{ij}$ is said to follow a Poisson distribution with parameter $\theta_{ij} > 0$, with notation $N_{ij}\mid \theta_{ij} \sim \text{Poisson}(\theta_{ij})$ if $N_{ij}$ has the following probability mass function:
\begin{equation} \label{eqn:poisson}
    P(N_{ij} = x \mid \theta_{ij}) = \frac{\theta_{ij}^x}{x!}\exp(-\theta_{ij}), \ x=1,2,\dots.
\end{equation}

\begin{table}[ht]
\centering
\caption{SRS contingency table}\label{table:SRS-frequency-table}
\begin{tabular}{c|cccc|c}
    \hline
       & Drug-$1$ & Drug-$2$ & $\cdots$ & Drug-$J$ & Total \\
    \hline
    AE-1 & $N_{11}$ & $N_{12}$ & $\cdots$ & $N_{1J}$  & $N_{1\bullet}$\\
    AE-2 & $N_{21}$ & $N_{22}$ & $\cdots$ & $N_{2J}$ & $N_{2\bullet}$\\
    $\vdots$ & $\vdots$ & $\vdots$ & $\ddots$ & $\vdots$ & $\vdots$\\
    AE-$I$ & $N_{I1}$ & $N_{I2}$ & $\cdots$ & $N_{IJ}$ & $N_{I \bullet}$\\\hline
    Total & $N_{\bullet 1}$ & $N_{\bullet 2}$ & $\cdots$ & $N_{\bullet J}$ & $N_{\bullet \bullet}$\\\hline
\end{tabular}
\end{table}

In many SRS data mining methods that are based on the Poisson model \eqref{eqn:poisson}, the parameter $\theta_{ij}$ is decomposed as $\theta_{ij} = \lambda_{ij}E_{ij}$, where  $\lambda_{ij}$ represents the relative reporting ratio, or \textit{signal strength}, measuring the expected ratio of the actual observed count ($N_{ij}$) to the null baseline expected count ($E_{ij}$) for the $i$-th AE and $j$-th drug.

To account for zero-inflation in SRS data, some methods employ the zero-inflated Poisson (ZIP) model for entries in the SRS contingency table. Given the binary 0/1 structural zero indicator $z \sim \text{Bernoulli}(p_0)$, the cell count $N_{ij}$ is said to follow a ZIP distribution with parameter $\theta_{ij} > 0$, if the conditional probability mass function of $N_{ij}$ given $z$, denoted $P(N_{ij} = x \mid \theta_{ij}, z)$, has the following form:

\begin{equation} 
        P(N_{ij} = x \mid \theta_{ij},z) = \begin{cases} 
      \frac{\theta_{ij}^x}{x!}\exp(-\theta_{ij}) & \text {if }\ z=0 \\
      \delta_0(x) &\text {if }\ z=1 
   \end{cases}, \quad x=1,2,\dots,
\end{equation}
where $\delta_0$ denotes the degenerate distribution concentrated at 0, i.e., $\delta_0(x)=\mathbb{I}\{x=0\}$. The marginal probability mass function of the ZIP model is:
\begin{equation} \label{zip_poisson}
        P_{\text{ZIP}}(N_{ij} = x \mid \theta_{ij},p_0) = \begin{cases} 
              p_0 + (1-p_0)\exp(-\theta_{ij}) &\text {if }\ x=0 \\
      (1-p_0)\frac{\theta_{ij}^x}{x!}\exp(-\theta_{ij}) & \text {if }\ x=1,2,3,\dots
   \end{cases}
\end{equation}

Some nonparametric Bayesian methods based on the Poisson model \eqref{eqn:poisson} can accommodate excess zeros without an explicit ZIP model over $N_{ij}$, instead using a flexible mixture-type prior distribution over $\lambda_{ij}$ (and consequently over $\theta_{ij}$) that assigns positive mass around (a small neighborhood of) zero. Examples include the hierarchical Dirichlet process (HDP) method \cite{hu2015signal} and nonparametric empirical Bayes models \cite{tan2025flex} which we will discuss later in this section

\begin{table}[ht]
\centering
\caption{Collapsed $2 \times 2$ contingency table for AE-$i$ and drug-$j$, $ \ i=1,\dots, I, \ j = 1, \dots, J$. $N_{\bullet \bullet}$ is the grand total; $N_{i \bullet}$ and $N_{\bullet j}$ are the $i$-th row marginal count and $j$-th column marginal count from the SRS frequency table (see Table \ref{table:SRS-frequency-table}).}\label{table:collapsed-contin-table}
\begin{tabular}{lll}
\hline
          & drug-$j$ & other drugs \\ \hline
AE-$i$      & $N_{ij}$      & $N_{i\bullet} - N_{ij}$          \\
other AEs & $N_{\bullet j}- N_{ij}$      & $N_{\bullet \bullet} - N_{i\bullet} - N_{\bullet j} + N_{ij}$           \\ \hline
\end{tabular}
\end{table}

\subsection{Proportional Reporting Ratio (PRR) and Reporting Odds Ratio (ROR)}

PRR and ROR are natural measures of factor association for a collapsed $2 \times 2$ table. Given the $2 \times 2$ contingency table for $i$-th AE and $j$-th drug displayed in Table \ref{table:collapsed-contin-table}, PRR and ROR are given by:

\begin{align*}
    &\text{PRR}_{ij} = \frac{\text{P}(\text{drug-$j$}\mid \text{AE-$i$})}{\text{P}(\text{drug-$j$}\mid \text{other AEs})}, \\
    &\text{ROR}_{ij} = \frac{\text{P}(\text{drug-$j$}\mid \text{AE-$i$})\ / \ \text{P}(\text{other drugs}\mid \text{AE-$i$})}{\text{P}(\text{drug-$j$}\mid \text{other AEs})\ / \ \text{P}(\text{other drugs}\mid \text{other AEs})}.
\end{align*}

PRR and ROR were introduced to detect signals for pharmacovigilance by Evans et al.\cite{evans2001use} and Rothman et al.\cite{rothman2004reporting}, respectively. Their estimators are defined as

\begin{align*}
    &\widehat{\text{PRR}}_{ij} = \frac{N_{ij} / N_{i\bullet}}{(N_{\bullet j} - N_{ij})/ (N_{\bullet \bullet} - N_{i\bullet})},\\
    &\widehat{\text{ROR}}_{ij} = \frac{N_{ij} / (N_{i\bullet} - N_{ij})}{(N_{\bullet j} - N_{ij})/ (N_{\bullet \bullet} - N_{i\bullet} - N_{\bullet j} + N_{ij})}.
\end{align*}

with approximated 95\% confidence intervals
\begin{align*}
    &\exp\left\{ \log(\widehat{\text{PRR}}_{ij}) \ \pm \ 1.96\sqrt{\frac{1}{N_{ij}} - \frac{1}{N_{i \bullet}} + \frac{1}{N_{\bullet j} - N_{ij}} - \frac{1}{N_{\bullet \bullet} - N_{i\bullet}}}   \right\},\\
    &\exp\left\{ \log(\widehat{\text{ROR}}_{ij}) \ \pm \ 1.96\sqrt{\frac{1}{N_{ij}} - \frac{1}{N_{i \bullet} - N_{ij}} + \frac{1}{N_{\bullet j} - N_{ij}} - \frac{1}{N_{\bullet \bullet} - N_{i\bullet} - N_{\bullet j} + N_{ij}}}   \right\}.
\end{align*}

By construction, PRR and ROR treat each AE-drug combination separately, and their uncertainty quantification depends on asymptotic approximations. These can lead to an inflated false discovery rate (FDR). FDR-adjusted variants of PRR and ROR have been proposed in the literature \cite{ding2020evaluation,Ahmed2010fdrror}. Ding et al.\ \cite{ding2020evaluation} evaluated the performance of PRR, ROR, and their FDR-adjusted version via simulation.

\subsection{Likelihood Ratio Test (LRT) based Methods}

Likelihood ratio test (LRT) approaches use a frequentist framework of null hypothesis significance testing for formally assessing the significance of signal strength parameters under parametric assumptions on the observed report counts (e.g., Poisson or zero-inflated Poisson model, see equations \eqref{eqn:poisson} and \eqref{zip_poisson}). They all derive an appropriate (likelihood ratio) test statistic and associated optimal cutoffs that strongly control frequentist type-I error and FDR, while ensuring sufficient power and reasonable sensitivity, owing to the optimality of likelihood ratio tests under a wide range of setups. Instead of relying on asymptotic approximations (e.g., as done in PRR, ROR, and one version of BCPNN), current approaches employ computational methods, viz., bootstrap and Monte Carlo sampling, for more accurate approximations of null sampling distributions. Below we review some of the prominent LRT based approaches for pharmacovigilance.

\subsubsection{Original LRT method}

The original LRT was introduced by Huang et al. (2011) \cite{huang2011likelihood}. For each drug-$j$, this method consider a Poisson model: $N_{ij} \sim \text{Poisson}(N_{i\bullet}\times p_{ij})$, where $p_{ij}$ represents the reporting rate of $j$-th drug for $i$-th AE; and $(N_{\bullet j} - N_{ij}) \sim \text{Poisson}((N_{\bullet \bullet} - N_{i\bullet}) \times q_{ij})$, where $q_{ij}$ represents the reporting rate of $j$-th drug for other AEs combined excluding the $i$-th AE. The LRT tests:
\begin{align*}
    &H_0: \ p_{ij} = q_{ij} = p_{0j} \quad  i=1,\dots, I; \\
    &H_{a,ij}: p_{ij}\neq q_{ij} \quad \text{for at least one $i$ among all I AEs.}\ 
\end{align*}
Under $H_0$, the maximum likelihood estimator (MLE) for $p_{0j}$ and $E_{ij}$ are $\hat p_{0j} = N_{\bullet j}/N_{\bullet \bullet}$ and $\hat E_{ij} = N_{i\bullet}N_{\bullet j}/N_{\bullet \bullet}$. Under $H_a$, the MLEs for $p_{ij}$ and $q_{ij}$ are $\hat p_{ij} = N_{ij}/N_{i\bullet}$ and $\hat q_{ij} = (N_{\bullet j} - N_{ij})/(N_{\bullet \bullet} - N_{i\bullet})$, respectively, leading to the likelihood ratio test statistic $\text{LR}_{ij} = {L_a (\hat p_{ij}, \hat q_{ij})}/{L_0(\hat p_{0j})}$. The consequent maximum likelihood ratio test statistic $\text{MLR}_j$ for drug $j$ and across all AEs $\{i\}$ is given by: 
\begin{align*}
    \text{MLR}_j = \max_i(\text{LR}_{ij}) &= \max_i\left( \frac{L_a (\hat p_{ij}, \hat q_{ij})}{L_0(\hat p_{0j})}\right)\\
    &= \max_i \left( \frac{(\hat{p}_{ij})^{N_{ij}}(\hat q_{ij})^{N_{\bullet j} - N_{ij}}}{(\hat p_{0j})^{N_{\bullet j}}}\right), \ i=1,\dots,I.
\end{align*}
For the one-sided LRT test, the test statistic is obtained by using $\text{LR}_{ij}\mathbbm{1}(\hat p_{ij} > \hat q_{ij})$ instead of $\text{LR}_{ij}$. There is no close form null distribution of the test statistic $\text{MLR}_j$, but it can be evaluated with a Monte Carlo approximation. Under $H_0$, given drug $j$, the AE-drug counts $\{N_{1j}, \dots, N_{Ij} \}$ conditioned on the row (AE) marginals $\{ N_{1\bullet}, \dots, N_{I\bullet}\}$ and $j$-th column (drug) marginal $N_{\bullet j}$ follow a multinomial distribution:
\begin{equation}\label{conditional_multinomial}
    (N_{ij},\dots, N_{Ij})\mid N_{\bullet j}; N_{1\bullet}, \dots, N_{I\bullet} \sim \text{Multinomial}\left( N_{\bullet j}, \left( \frac{N_{1\bullet}}{N_{\bullet \bullet}}, \dots, \frac{N_{I\bullet}}{N_{\bullet \bullet}}\right) \right).
\end{equation}

The distribution of $\text{MLR}_j$ is evaluated by the Monte Carlo replications drawn from the above multinomial model. 

Ding et al. \cite{ding2020evaluation} later adapted the original LRT method to both AE-based and drug-based formulations, showing improved performance compared to some other signal detection methods. Huang et al.\cite{huang2017zero} proposed a zero-inflated Poisson model \eqref{zip_poisson}-based likelihood ratio test (ZIP-LRT) method to handle the excess zeros in SRS data.

\subsubsection{Extended Likelihood Ratio Test (Ext-LRT)}

Zhao et al. \cite{zhao2018extended} extended the original LRT approach to detect signals of AEs simultaneously for multiple drugs $\{j\}$. This is often the case when a specific group of drugs is of interest. Suppose the drug class of interest includes $M$ different drugs. With the same model assumptions as the original LRT, these drugs lead to a group of LRT tests with null hypotheses: $H_{0m}\ : \ p_{im} = q_{im} = p_{0m}$ for all $i = 1,\dots, I; \ m = 1,\dots, M$. Then the test statistic for testing $H_{0m}$ versus $H_{am}$ is
\begin{align*}
        \text{MLR}_m = \max_i(\text{LR}_{im}) &= \max_i\left( \frac{L_a (\hat p_{im}, \hat q_{im})}{L_0(\hat p_{0m})}\right)\\
    &= \max_i \left( \left( \frac{N_{im}}{E_{im}}\right)^{N_{im}} \left( \frac{N_{\bullet m} - N_{im}}{N_{\bullet m} - E_{im}}\right)^{N_{\bullet m} - N_{im}}\right), \ i=1,\dots,I; \ m = 1,\dots, M.
\end{align*}
The global null hypothesis for detecting any signals among the drugs of interest is $H_0\ :\ \underset{m}{\cap} \underset{i}{\cap}\{ p_{im} = q_{im} = p_{0m} \} $ versus the global alternative hypothesis $H_a \ : \  \underset{m}{\cup} \underset{i}{\cup}\{p_{im} > q_{im}\}$. The corresponding test statistic is given by:
\[
\text{MLR}^{1:M} = \max_m(\text{MLR}_m) = \max_m(\max_i (\text{LR}_{im}\mathbbm{1}(\hat p_{im}> \hat q_{im}))),
\]
where $i = 1,\dots, I$, $m = 1,\dots, M$. Similar to the original LRT, the distribution of the test statistic under the null hypothesis is evaluated by the Monte Carlo approximation from the conditional Multinomial distribution \eqref{conditional_multinomial}. The Extended LRT (Ext-LRT) considers an entire drug class, usually defined as a group of drugs with similar chemical structures or functions. This approach enables simultaneous signal detection across the class and effectively controls type I error and FDR while maintaining good power and sensitivity for data with no zero-inflation. When zero-inflation is present, an extended version of the likelihood ratio test (Ext-ZIP-LRT) can be used as proposed in Zhao et al. \cite{zhao2018extended}, which extends the Ext-LRT to handle a zero-inflated Poisson model for $N_{ij}$.

\subsubsection{Pseudo-LRT approach}\label{sec:pseudo-lrt}

The original LRT \cite{huang2011likelihood} method and its extensions discussed above assume for a specific drug $j^*$ and AE $i$, $N_{ij^*} \sim \text{Poisson}(N_{i\cdot} \times p_{ij^*})$ and $(N_{\cdot j^*} - N_{ij^*}) \sim \text{Poisson}((N_{\cdot \cdot} - N_{i\cdot})\times  q_{ij^*})$ or the corresponding zero-inflated Poisson model version, and perform signal detection by comparing the evidence for the null hypothesis with the alternative. This parametrization of LRT assumes two separate Poisson/zero-inflated Poisson models for each $2\times 2$ contingency table that may lead to model violation when all $2\times 2$ tables are considered simultaneously; e.g., the implied model for the marginal counts $N_{\bullet j^*}$ from all $2\times 2$ may tables may considerably violate the model assumption $(N_{\cdot j^*} - N_{ij^*}) \sim \text{Poisson}((N_{\cdot \cdot} - N_{i\cdot})\times  q_{ij^*})$.  

Chakraborty et al. \cite{chakraborty2022use} considered an alternative relative reporting rate-based parametrization and develop a pseudo-LRT approach. For a specific drug $j^*$ and AE $i$, this method assumes that $N_{ij^*} \sim \text{Poisson}(\lambda_{ij^*}\times E_{ij^*})$ or $N_{ij^*} \sim \text{ZIP}(\lambda_{ij^*}\times E_{ij^*}, p_0)$. Under $H_0$, $p_0$ is estimated using profile likelihood maximization. The pseudo-LRT method has been shown to improve computational efficiency and enhance power in the presence of zero-inflation through simulation experiments. 

\subsection{Bayesian and empirical Bayes methods}

Bayesian and empirical Bayes methods have become an important class of approaches in SRS data mining for pharmacovigilance. Let $g = g(\lambda | \theta)$ be a common prior density for the signal strength parameter ($\lambda>0$) indexed by a vector of hyperparameters $\theta$. Combining the prior $g$ with the observed SRS data likelihood (e.g., Poisson model):

\[
N_{ij} \mid \lambda_{ij} \sim \text{Poisson}(\lambda_{ij}E_{ij}),
\]
using the Bayes theorem yields the posterior distribution for $\lambda_{ij}$. Under an empirical Bayes framework, the hyperparameters $\theta$ in the prior distribution $g$ are estimated by maximizing the marginal likelihood:
\[
\hat{\theta} = \argmax \prod_{i=1}^I\prod_{j=1}^J\int_0^{\infty}g(\lambda_{ij}\mid \theta)f_{\text{Poisson}}(N_{ij}\mid \lambda_{ij} )d\lambda_{ij}.
\]
The estimated posterior distribution for $\lambda_{ij}$ is then obtained by plugging in the estimated prior hyperparameters. These methods yield posterior/estimated posterior distributions that naturally incorporate uncertainty and enable shrinkage estimation. A gamma distribution-based prior is a common choice under a Poisson likelihood due to conjugacy. In this subsection, we review several Bayesian models based on a gamma or a gamma-mixture prior. The gamma distribution with parameter (shape= $\alpha$, rate= $\beta$) has the following probability density function:
\[ 
f_{\text{gamma}}(x\mid \alpha,\beta) = \frac{\beta^{\alpha}}{\Gamma(\alpha)}x^{\alpha-1}\exp^{-\beta x}, \ x>0.
\]

The resulting shrinkage in the posterior of $\lambda_{ij}$ is particularly important in pharmacovigilance for stabilizing noisy estimation, as inference with SRS data which can involve high-dimensional and sparse signal-strength parameters, with most AE-drug pairs being non-signals and only a handful being signals. To illustrate the impact of the shrinkage effect,  consider an AE-drug pair with an observed count of $N = 1$ and an expected null baseline count of $E = 0.01$. The naive estimator of signal strength from the Poisson model \eqref{eqn:poisson} is $\hat \lambda = N/E = 100$. However, under a simple Bayesian model with a prior of $\lambda \sim \text{Gamma}(0.5, 0.5)$, the posterior mean of the signal strength $\lambda$ for this pair is shrunk toward the prior mean: $E(\lambda \mid N) = (1 + 0.5) / (0.01 + 0.5) \approx 2.94$. This substantially avoids noisy posterior estimates by borrowing information from the prior distribution.

Depending on the choice of prior, Bayesian analyses can range from simple parametric models, such as a single gamma prior, to more flexible semi-parametric and nonparametric formulations that account for heterogeneity across signals, non-signals, and structural zeros. As a result, Bayesian methods offer interpretable posterior probabilities of being a signal, richer assessments of AE relevance, and principled uncertainty quantification. Specifically, for each AE-drug pair of interest, a posterior probability of being a signal
\begin{equation}\label{eqn:signal-prob}
    P(\lambda_{ij} > 1 + \epsilon\mid N_{ij}),
\end{equation}
for some small $\epsilon$ (e.g., $\epsilon = 0.001$) is evaluated and compared against a cutoff such as 0.95 to determine whether the corresponding AE-drug pair is a detect signal. Alternatively, Tan et al. \cite{tan2025flex} suggest visualizing the posterior estimation of the signal strength parameter $\lambda_{ij}$ using the posterior density or posterior median along with the corresponding 90\% credible interval. The signal strength estimation performance of major Bayesian/empirical Bayes methods was systematically evaluated by Tan et al. \cite{tan2025flex} through comprehensive simulations. The results demonstrated that nonparametric Bayesian and empirical Bayes approaches provide superior performance in estimating signal strength $(\lambda_{ij})$ compared with traditional Bayesian models, while achieving comparable results in signal detection metrics (i.e., FDR and sensitivity).

The main line of Bayesian models for analyzing SRS frequency table is built upon the use of a gamma/gamma mixture prior. The simplest model uses a single gamma prior, which leads to closed-form shrinkage posterior signal strength parameter $\lambda_{ij}$ estimates. The Gamma-Poisson Shrinker (GPS)\cite{dumouchel1999bayesian} method uses a two-gamma mixture prior that distinguishes signal/non-signal components in SRS data, allowing more flexible modeling, compared with the single gamma model. More recent developments have focused on nonparametric Bayesian/empirical Bayes frameworks such as the hierarchical Dirichlet process (HDP)\cite{hu2015signal} method, Koenker and Mizera (KM)\cite{koenker2014convex} approach, Efron\cite{efron2016empirical} approach, and general-gamma\cite{tan2025flex} method, which further relax parametric assumptions and obtain Bayesian inference of signal strength parameter $\lambda_{ij}$'s through a data-driven way. In the following subsections, we provide detailed descriptions of these methods.

\subsubsection{Bayesian Confidence Propagation Neural Network (BCPNN)}

The Bayesian Confidence Propagating Neural Network (BCPNN), introduced by Bate et al. \cite{bate1998bayesian}, is one of the earliest Bayesian approaches for signal detection in pharmacovigilance. Unlike most of the Bayesian models discussed in this section, which use a Poisson model for each SRS contingency table cell count, BCPNN is formulated using a binomial model for each cell count (which can be viewed as a consequence of independent Poisson cell counts conditional on the row and column totals). Let $p_{ij}$ be the probability of the occurrence of the $(i, j)$-th cell, and let $p_{\bullet j}$ and $p_{i\bullet}$ be the marginal column and row probabilities, each with a Uniform(0, 1) prior distribution. The hierarchical model assumptions for the cell counts and marginal totals are: $N_{ij} \mid p_{ij} \sim \text{Bin}(N_{\bullet\bullet}, p_{ij})$, $N_{i\bullet} \mid p_{i\bullet} \sim \text{Bin}(N_{\bullet\bullet}, p_{i\bullet})$, $N_{\bullet j} \mid p_{\bullet j} \sim \text{Bin}(N_{\bullet\bullet}, p_{\bullet j})$; with the prior $p_{ij} \sim \text{Beta}(1,\beta_{ij})$. The prior parameters $\{\beta_{ij}\}$ are determined/estimated by setting the prior mean of $p_{ij}$ equal to the product of the posterior means of $p_{i\bullet}$ and $p_{\bullet j}$.

Different from Gamma-Poisson Bayesian models that directly target the signal strengths ($\lambda_{ij}$'s), BCPNN builds upon information theory by quantifying the association between a drug and an adverse event (AE) through the information component (IC), a measure derived from the mutual information between the marginal and joint probabilities of AE-drug co-occurrence:
\[\text{IC}_{ij} = \log_2 \frac{p_{ij}}{p_{i\bullet} p_{\bullet j}}, \ i=1,\dots, I, \ j = 1,\dots, J.\]
The Bayesian inferences on $\{\text{IC}_{ij}\}$ are based on an asymptotic normal approximation of their posterior distributions, characterized by the estimated asymptotic mean and variance as follows\cite{bate1998bayesian}:
$$\hat{\text{E}}(\text{IC}_{ij}) \approx \log_2 \frac{(N_{ij}+1)(N_{\bullet\bullet}+2)^2}{(N_{\bullet\bullet}+\hat{\beta}_{ij})(N_{i\bullet}+1)(N_{\bullet j} +1)},$$
$$\hat{\text{Var}}(\text{IC}_{ij}) \approx \frac{1}{(\log 2)^2}\left(\frac{N_{\bullet\bullet}-N_{ij}+\hat{\beta}_{ij} -1}{(N_{ij}+1)(1+N_{\bullet\bullet}+\hat{\beta}_{ij})} + \frac{N_{\bullet\bullet}-N_{i\bullet}+1}{(N_{i\bullet}+1)(N_{\bullet\bullet}+3)}+\frac{N_{\bullet\bullet}-N_{\bullet j}+1}{(N_{\bullet j}+1)(N_{\bullet\bullet}+3)}\right).$$
The posterior inference of BCPNN does not account for the joint variability among all $O/E$ values, which consequently leads to the notable limitation that this method may fail to control the FDR. A posterior probability-based FDR adjustment can be applied to BCPNN to improve signal detection performance \cite{muller2006fdr}.

\subsubsection{Gamma-Poisson Shrinker (GPS) and Multi-item Gamma-Poisson Shrinker (MGPS)}\label{sec:gps}

The Gamma Poisson Shrinker (GPS) \cite{dumouchel1999bayesian} and its extension, Multi-item Gamma-Poisson Shrinker (MGPS) \cite{dumouchel2001empirical},  are among the most prominent Bayesian approaches for pharmacovigilance signal detection. The both methods use a semi-parametric mixture of two gamma distribution components as prior (we call it ``2-gamma'' prior in this paper). The prior density function is given by:
\[
g(\lambda \mid \alpha_1, \beta_1, \alpha_2, \beta_2, \omega) = \omega f_{\operatorname{gamma}}(\lambda \mid \alpha_1, \beta_1) + (1-\omega)f_{\operatorname{gamma}}(\lambda \mid \alpha_2, \beta_2), 
\]
where $\alpha_1, \alpha_2, \beta_1, \beta_2 > 0$ are the mixture component-specific prior parameters, $\omega \in [0, 1]$ represents the weight of the mixture component in the prior distribution, and $f_{\operatorname{gamma}}(\cdot \mid \alpha,\beta)$ is the probability density function of a gamma distribution with parameters $\alpha$ and $\beta$. In SRS databases, the majority of AE-drug pairs are non-signals, with their underlying signal strength $\lambda_{ij}$ typically concentrated around 1. We commonly have a much smaller number of AE-drug pairs that correspond to true signals showing strong AE-drug association with $\lambda_{ij} > 1$. This pattern is well accommodated by a two-component gamma mixture prior: one component, with a large mixing weight, captures the non-signal AE-drug pairs with signal strength $\lambda_{ij} \approx 1$, while the other mixture component represents the relatively rare signal AE-drug pairs with $\lambda_{ij}>1$. However, this prior distribution setting does not account for inflated zeros and substantial heterogeneity in the $\lambda_{ij}$ values within each component, which are common in the SRS data. Tan et al. \cite{tan2025flex} demonstrated this limitation through simulations. Consequently, we recommend a preliminary assessment of zero-inflation before applying GPS or MGPS for signal detection in SRS data.

DuMouchel (1999) \cite{dumouchel1999bayesian} proposed an empirical Bayes approach for the 2-gamma model, which maximizes the marginal likelihood of the observed counts $\{N_{ij}\}$. This provide maximum marginal likelihood estimates $\{\hat \alpha_1, \hat \beta_1, \hat \alpha_2, \hat \beta_2, \hat \omega\}$ for the prior hyperparameters. Given these estimates, the estimated posterior distribution of $\lambda_{ij}$ conditional on $N_{ij}$ can be derived by the Bayes rule, which retains the same two-component gamma mixture form and is used for signal detection.

Based on the Poisson model \eqref{eqn:poisson}, the marginal probability mass function of $N_{ij}$ is a mixture of two negative binomial distributions:
\[
p(N_{ij} = n_{ij} \mid \alpha_1, \beta_1, \alpha_2, \beta_2, \omega) 
= \omega f_{\text{NB}}(n_{ij} \mid \alpha_1,\beta_1, E_{ij}) + (1-\omega) f_{\text{NB}}(n_{ij} \mid \alpha_2,\beta_2, E_{ij}),
\]
where $f_{\text{NB}}(n \mid \alpha,\beta, E) = \frac{\Gamma(n+\alpha)}{\Gamma(\alpha)n!}\left( \frac{\beta}{E+\beta} \right) ^{n} \left(\frac{E}{E+\beta} \right) ^{\alpha}$; $n=1, 2, \dots$, is the negative binomial probability mass function. The parameters $(\alpha_1, \beta_1, \alpha_2, \beta_2, \omega)$ are estimated by maximizing their marginal likelihood:
\[
(\hat \alpha_1, \hat \beta_1, \hat \alpha_2, \hat \beta_2, \hat \omega) = \argmax_{\alpha_1, \beta_1, \alpha_2, \beta_2, \omega} \prod_{i}\prod_j p(N_{ij} = n_{ij} \mid \alpha_1, \beta_1, \alpha_2, \beta_2, \omega).
\]
Since closed-form expressions for $(\hat \alpha_1, \hat \beta_1, \hat \alpha_2, \hat \beta_2, \hat \omega)$ do not exist, the marginal likelihood is maximized using numerical optimization methods. Currently, implementation of the GPS model is available in R packages: ``PhViD'', ``openEBGM'' and ``pvEBayes''.

\subsubsection{Hierarchical Dirichlet Process (HDP) method}

To address the complexity of the SRS data, nonparametric methods have been adopted. The Hierarchical Dirichlet Process methods (HDP), introduced by Hu et al. \cite{hu2015signal}, extend the 2 component gamma mixture model in GPS by adopting a fully nonparametric Bayesian prior for the signal strength $\lambda_{ij}$. Instead of a fixed number of mixture components in the prior distribution, the HDP induces a Dirichlet process prior, which is an infinite mixture representation for $\lambda_{ij}$. Each component corresponds to a degenerate distribution with an atom drawn from a baseline Gamma distribution. The associated mixing weights decrease in expectation. This allows the model to capture multiple subgroups for $\lambda_{ij}$ values---including within the signal, non-signal, and structural-zero groups---while adaptively maintaining a relatively small number of effective or non-empty components. The model is described as follows. For each drug $j$, the signal parameters are modeled as
\begin{gather*}
    \lambda_{ij} \sim G_j \sim \operatorname{DP}(\rho_j, G_{0j}), \\
    G_{0j} \equiv \operatorname{Gamma}(\alpha_j, \alpha_j), \\
    \alpha_j \sim \operatorname{Uniform}(0,1), \\
    \rho_j\sim \operatorname{Uniform}(0.2,10),
\end{gather*}
where $\operatorname{DP}(\rho_j, G_{0j})$ denotes a Dirichlet process with precision parameter $\rho_j$ and baseline distribution $G_{0j}$ which is a unit-mean Gamma distribution of the form $\operatorname{Gamma}(\alpha_j, \alpha_j)$. Hu et al. \cite{hu2015signal} assign flexible uniform hyperprior distributions to $\alpha_j$ and $\rho_j$ to ensure the adaptability of the resulting full Bayesian model. In practice, the infinite mixture is approximated by a truncated finite mixture using the stick-breaking representation of the Dirichlet process. The implementation of the HDP model heavily relies on Markov Chain Monte Carlo (MCMC) methods, which can be computationally intense. 

\subsubsection{The Koenker and Mizera (KM) approach}

Koenker and Mizera \cite{koenker2014convex} proposed a nonparametric empirical Bayes model. It has been introduced and adapted for pharmacovigilance by Tan et al. \cite{tan2025flex}. Unlike parametric or semi-parametric Bayesian methods that require a specific prior form for the signal strength parameters $\lambda_{ij}$, the KM model assumes a finite discrete prior $g$ for $\{\lambda_{ij}\}$ with support of size $K$, $\lambda \in \{ v_1,...,v_K\}$, $K<\infty$ and associated prior probability masses: $\{ g_1,...,g_K\}$ with $g_k \geq 0$; $\sum_{k=1}^K g_k = 1$. Under this prior assumption, the marginal probability mass for the $(i,j)$-th observation is obtained as:
\[
f_{ij} = \sum_{k=1}^K f_{\text{pois}}(N_{ij} \mid  v_k E_{ij}) \times g_k.
\]
This constructs a nonparametric empirical Bayes method for pharmacovigilance. For empirical Bayes estimation, Koenker and Mizera \cite{koenker2014convex} suggest estimating the prior parameters $\{ g_1, \dots, g_K \}$ through the following constrained maximization problem:
\[
\max \left\{ \sum_{i=1}^I\sum_{j=1}^J\log(f_{ij}): \quad f_{ij} = \sum_{k=1}^K f_{\text{pois}}(N_{ij}\mid  v_k E_{ij})\times g_k,\quad g_k \geq 0, \quad \sum_{k=1}^K g_k=1\right \},
\] 
We note that the KM method is highly flexible, but its performance depends on the $\lambda$-grid values $\{ v_1, \ldots, v_K \}$, which are not straightforward to determine a priori. Tan et al. \cite{tan2025flex} proposed a histogram-based grid selection approach. The implementation of the KM method is available from the ``REBayes'' and ``pvEBayes'' packages.

\subsubsection{Efron approach}

As an alternative to the KM model, Efron’s approach \cite{efron2016empirical} also provides a discrete, nonparametric empirical Bayes model for pharmacovigilance. This model also assumes a finite discrete support of size $K$,  $\lambda \in \{ v_1,...,v_K\}$, $K<\infty$. However, instead of estimating the prior probabilities directly, the mass on the grid has an exponential family form of density:
\[
g = g(\alpha) = \exp\{ Q \alpha - \phi(\alpha) \},
\]
where, $\alpha$ is a $p$-dimensional parameter vector, $Q$ is a known $K \times p$ structure matrix, and $\phi(\alpha)$ is a normalizing constant ensuring that $g$ is a proper mass function. A common choice for $Q$, suggested by Efron \cite{efron2016empirical}, is a natural spline basis over $\{ v_1,...,v_K\}$with $p$ degrees of freedom. Then, the prior parameter $\alpha$ is estimated through a ridge-penalized marginal likelihood:
\[
\log L_{E}(\alpha ; c_0, p) = \left\{ \sum_{i=1}^I\sum_{j=1}^J \log P_{ij}^{\top} g(\alpha) - c_0\left( \sum_{l=1}^p \alpha_l^2\right)^{1/2} \right\},
\]
with $P_{ij} = (f_{\text{pois}}(N_{ij} \mid v_k E_{ij}): k= 1, \dots, K)$; and pre-specified $c_0 > 0$ and $p$. This optimization can be solved using standard gradient-based algorithms; Efron specifically recommends a Fisher-scoring type procedure \cite{efron2016empirical}. Tan et al. \cite{tan2025flex} suggest an Akaike information criterion (AIC) based hyperparameter $(c_0, p)$ selection. Let $H_E(\alpha ; c_0, p)$ denote the hessian matrices corresponding to $\log L_{E}(\alpha ; c_0, p)$. The AIC for an Efron model with hyper parameters $(c_0, p)$  is defined as:
\begin{equation*}
    AIC_{\text{E}}(c_0, p) = 2\times \text{trace}(\text{F}) - 2\log L_{E;0}(\hat \alpha),
\end{equation*}
where $\text{F} = H_E^{-1}(\hat \alpha; c_0, p)H_{E}(\hat \alpha ;0,  p)$ is the degrees of freedom matrix (Wood, 2017 \cite{wood2017generalized}) associated with the penalized marginal likelihood.

\subsubsection{General-Gamma method} \label{general-gamma}

If there exists zero-inflation among the observed counts, for example, due to structural zero AE-drug pairs, this can be explicitly accommodated through a straightforward extension of the 2-gamma prior model (see section \ref{sec:gps}). This is achieved by introducing an additional ``zero-inflation'' component $\lambda_{ij}$ to accommodate small $\lambda_{ij}$ values: 
\begin{align*}
    g(\lambda \mid \alpha_1, \beta_1, \alpha_2, \beta_2, \alpha_3, \beta_3,\omega_1, \omega_2) = & \omega_1 f_{\operatorname{gamma}}(\lambda \mid \alpha_1, \beta_1) + \omega_2f_{\operatorname{gamma}}(\lambda \mid \alpha_2, \beta_2) \\
    & + (1-\omega_1 - \omega_2)f_{\operatorname{gamma}}(\lambda \mid \alpha_3, \beta_3).
\end{align*}
Additionally, the existence of heterogeneity of signal AE-drug combinations naturally leads to a $K$ component gamma mixture model:
\begin{equation}\label{eqn:K-gamma-mixture}
    g(\lambda\mid R, H, \Omega) = \sum_{k=1}^K \omega_k f_{\text{gamma}} \left(\lambda\mid \alpha = r_k, \beta = \frac{1}{h_k}\right), 
\end{equation}
where $\Omega = \{\omega_1, \dots, \omega_K\}$, $R = \{r_1, \dots, r_K\}$, and $H = \{h_1, \dots, h_K\}$ are component-specific parameters. However, determining the number of mixture components $K$ for this model is challenging. To address this, Tan et al. \cite{tan2025flex} proposed a nonparametric empirical Bayes approach (general-gamma) which adapts the sparse mixture model\cite{fruhwirth2019here, malsiner2016model, malsiner2017identifying} by assuming mixture weight with a Dirichlet hyper prior. This method improves GPS and its gamma mixture extensions by adaptively determining the effective number of mixture components $K$ from the data.  Specifically, the model begins with an overfitted gamma mixture model (see equation \ref{eqn:K-gamma-mixture}), with number of mixture component $K$ chosen to be large (e.g., $K=100$), where the mixture weights follow a Dirichlet hyper prior, $\Omega = \{\omega_1, \dots, \omega_K\} \sim \text{Dirichlet}(\alpha,\alpha, \dots, \alpha)$ with $\alpha < 1$.  The Dirichlet hyper prior leads to a sparse mixture by encouraging many of the mixture weights to shrink toward zero, thereby allowing the model to automatically identify and retain only the most important components. 

Choosing $\alpha \in (0,1)$ for the Dirichlet prior induces sparsity in the fitted mixture, such that many components become \textit{inactive} (with weights shrunk to zero), while the number of \textit{active} components (with positive weights) converges to the true number of distinct mixture components in the population. More precisely, under suitable regularity conditions, sparse overfitted mixtures are known to asymptotically converge to the ``true'' population mixture (Rousseau and Mengersen, 2011 \cite{rousseau2011asymptotic}) provided that the Dirichlet hyperparameter $\alpha$ is chosen smaller than $d/2$, where $d=2$ is the dimension of the component-specific parameter set $\{R, H\}$. 

The prior parameters $\{\Omega, R, H\}$ are estimated by maximizing the marginal likelihood through an efficient expectation conditional maximization (ECM). The general-gamma model is available in an R package ``pvEBayes''.

\section{Illustration}\label{sec:illustration}

FAERS, EudraVigilance, and VigiBase are three of the largest and most widely used pharmacovigilance databases in the world. In this section, we provide an overview of these systems from the perspective of obtaining SRS dataset for pharmacovigilance investigations. In what follows, we first offer a brief summarization of the pharmacovigilance databases with a recommended procedure for obtaining SRS datasets from VigiBase and EudraVigilance with public accessibility. After that, we present an illustrative example demonstrating the process of obtaining opioid-related data from FAERS and VigiBase, followed by downstream signal detection and signal strength estimation analyses using the general-gamma method and the pseudo-Likelihood Ratio Test (pseudo-LRT) approach. These two methods represent contemporary nonparametric empirical Bayes and likelihood based methods that are shown to be powerful for SRS data mining.

EudraVigilance is the European Union’s database for managing and analyzing reports of suspected adverse drug combinations. It was established and maintained by the European Medicines Agency (EMA) to support pharmacovigilance activities throughout the European Economic Area (EEA). It primarily collects two types of AE reports: (1) individual case safety reports (ICSR) submitted by national authorities, marketing authorisation holders, healthcare professionals, or patients through national systems; (2) Suspected unexpected serious adverse event reports from interventional studies conducted under the EU clinical trials regulation. The access to EudraVigilance data is categorized into various access levels from level 1 (limited to a public subset of aggregated ICSR data) to level 3 (full access to EudraVigilance). For Academia, obtaining data beyond the publicly available subset of ICSR data requires a formal application to the EMA with several requirements\cite{EMA2025EudraVigilance}. 

The FDA Adverse Event Reporting System (FAERS) is the United States Food and Drug Administration’s database for postmarketing safety surveillance. It serves as a core component of the FDA’s postmarketing safety surveillance program, enabling the detection of potential safety signals, risk assessment, and regulatory decision-making regarding approved medical products. It compiles AE reports submitted to the FDA by healthcare professionals, consumers, and manufacturers. Among all the drug safety databases, FAERS provides the highest level of accessibility by offering raw FAERS AE reports through quarterly data files on the FAERS website.

VigiBase is the global database of Individual Case Safety Reports maintained by the Uppsala Monitoring Centre (UMC) on behalf of the World Health Organization (WHO). It was established in 1968 as part of the WHO Programme for International Drug Monitoring (PIDM). VigiBase is the largest international repository of SRS reports, currently comprising over 40 million reports from more than 180 WHO PIDM members, including reports from FAERS, EudraVigilance, and other pharmacovigilance databases worldwide. Similar to EudraVigilance, the accessibility of VigiBase data is organized by different access levels corresponding to different stakeholders. Aggregated AE counts for a given drug can be obtained from the VigiAccess website. 

\begin{table}[ht]
\caption{Overview of major Pharmacovigilance Databases: EudraVigilance, VigiBase, and FAERS}
\label{tab:overview-database}
\centering
\begin{tabular}{llll}
\hline
              & EudraVigilance                  & VigiBase                            & FAERS                                   \\ \hline
Region        & European Union                  & Global                              & United States                           \\
Maintained by & European Medicines  & WHO Uppsala Monitoring  & U.S. Food and Drug  \\
& Agency (EMA) & Centre (UMC)& Administration (FDA)\\
Start year    & 2001                            & 1968                                & 1998                                    \\
Report type   & SRS and CT                      & SRS                                 & SRS                                     \\ 
Coding system &MedDRA &MedDRA &MedDRA \\
Publicly access &aggregrated AE-drug counts& aggregrated AE-drug counts& raw AE reports\\
level &&&\\\hline
\end{tabular}
\end{table}

Table~\ref{tab:overview-database} provides a comparison of key characteristics of EudraVigilance, VigiBase, and FAERS. All three pharmacovigilance databases use MedDRA (Medical Dictionary for Regulatory Activities) for coding and classifying adverse events (AEs), indications, and related medical information contained in AE reports. This reduces variability due to differences in language, local medical terminology, or reporting style.

In pharmacovigilance, the accessibility of spontaneous reporting system (SRS) data has important implications for data acquisition, processing, and downstream SRS data mining. The publicly available FAERS data allow researchers to directly construct complete SRS contingency tables for a specific drug of interest, with a well-defined reference group of ``other drugs''. This facilitates better estimation of expected null counts for both signal detection and signal strength estimation approaches. In contrast, the restricted accessibility of EudraVigilance and VigiBase for researchers, where only aggregated AE-drug counts can be retrieved, limits the ability to define a comparable reference comparison ``other drugs''.  Therefore, while pharmacovigilance databases, e.g., EudraVigilance and VigiBase,  provide broader geographic coverage, their data access policies impose substantial constraints on either immediate accessibility or time needed to acquire the data for downstream SRS data mining for pharmacovigilance investigation.

In practice, for researchers with access limited to publicly available data from EudraVigilance and VigiBase, we recommend the following procedure to obtain SRS data for SRS data mining.
\begin{itemize}
    \item[1] Specify the drug or a group of drugs of interest.

    \item[2] (optional) Identify a set of drugs considered independent of the drug(s) of interest to serve as reference drugs. Note that the drug(s) of interest should not be presented in the column of reference drugs.  

    \item[3] For each drug, obtain the corresponding AE-drug counts from VigiBase or the EudraVigilance website.

    \item[4] Tabulate data obtained in step 3 into a large SRS contingency table. The reference drug column is obtained by collapsing columns that correspond to drugs that are considered independent of the drug of interest.

    \item[5] (optional) Select a set of AE of interest and collapse other AE rows into a reference AE row ``other AEs''.
\end{itemize}

We note that although some of the SRS mining methods do not directly assume the existence of a reference drug column, and we label step 2 as optional, we still recommend obtaining the reference drug column for a better estimation of the expected null baseline counts. The inclusion of "other AEs" and "other drugs" (reference groups) is typically important for SRS data mining. Because SRS data lack a true population denominator of exposed cases, SRS data mining methods rely on estimated expected baseline count from the SRS data. Consequently, the "other" categories provide the necessary data mass for better estimation of the background reporting rate. Without a carefully constructed reference row and column, downstream SRS mining could be unreliable, resulting in an uncontrolled false discovery rate or inflated variance of posterior estimation of signal strength $\lambda$. In the next subsection, we provide an example of obtaining opioid SRS data from both FAERS and VigiBase.

\subsection{FAERS and VigiBase opioid dataset}\label{sec:opioid-datasets}

In this subsection, we follow the procedure suggested above to obtain opioid SRS datasets from FAERS and VigiBase. The drugs of interest included Codeine, Fentanyl, Oxycodone, Pentazocine, and Tramadol. For FAERS,  we first download the raw adverse event (AE) report data during 2014Q3-2024Q4 from the FDA’s official website.  For each quarterly dataset, cases in which any of the target opioids were identified as the primary suspect drug were extracted to form an SRS contingency table of AE-drug pairs. These quarterly tables were subsequently merged to produce the raw-opioid-FAERS dataset. To provide a reliable reference group, we identified the 1200 most frequently reported non-opioid drugs across AE reports in all quarters and collapsed them into a single reference category representing “all other drugs.”

For VigiBase, we used the software SurVigilance \cite{mukhopadhyay2025survigilanceapplicationaccessingglobal} to retrieve aggregated AE-drug count data for the same set of opioid drugs and for the 1200 reference drugs identified from FAERS. We note that due to the accessibility difference between VigiBase and FAERS, we are not able to filter the AE-drug count for the same period of time, making the data from VigiBase have a longer time range if they are introduced before 2014Q3. These data were combined to construct the raw-opioid-Vigi dataset. Finally, to focus on mental health-related adverse outcomes, we filtered the adverse event terms by mental health-related keywords (presented in Supplemental Table S1). AEs that contain or partially contain one of the keywords are selected as AE of interest; otherwise, they are categorized into the reference row "other AEs". The above procedure results in two SRS subsets: FAERS-opioid-mental and VigiBase-opioid-mental.

The constructed SRS dataset includes mental health-related AE counts for opioid drugs collected from FAERS and VigiBase. The VigiBase-opioid and FAERS-opioid datasets cover AE-drug counts across 6 drug categories (Codeine, Fentanyl, Oxycodone, Pentazocine, Tramadol, Other drugs)  and distinct AEs (100 in VigiBase and 243 in FAERS). The marginal count of AE reports for the drugs of interest, other drugs, and the total number of AE-drug reports in each opioid data is presented in Table \ref{table-marginal-count}. Most of the AE reports for opioid drugs are in these ten years, while the drugs in the reference column ``other drugs'' are defined as the most frequently appearing drugs (e.g., Aspirin,... ), and most of them have been introduced for a much longer period of time. We see a larger discrepancy in the AE counts for the reference drug column compared to opioid drugs. Both opioid datasets are available in Supplemental Section S1.1-S1.2 or can be accessed directly from the “pvEBayes” package in R.

The results of the SRS data mining of these two datasets with the general-gamma and pseudo-LRT methods are presented in the next subsection.

\begin{table}[ht]
\resizebox{\textwidth}{!}{%
\centering
\begin{tabular}{llllllll}
\hline
Dataset                & Codeine & Fentanyl & Oxycodone & Pentazocine & Tramadol &Other drugs & Grand total \\ \hline
VigiBase-opioid-mental & 46,162   & 509,596   & 665,954    & 8,133        & 487,125   & 77,140,702 &78,857,672    \\
FAERS-opioid-mental    & 29,668   & 132,880   & 573,421    & 1,537        & 138,253   & 28,718,599 &29,594,358    \\ \hline
\end{tabular}
}
\caption{AE-drug count for each drug and the grand total in VigiBase-opioid-mental and FAERS-opioid-mental datasets. }\label{table-marginal-count}
\end{table} 

\subsection{SRS Data mining with obtained datasets}\label{sec:disproportionality-analysis}

\begin{figure}[ht]
    \centering
    \includegraphics[width=\textwidth]{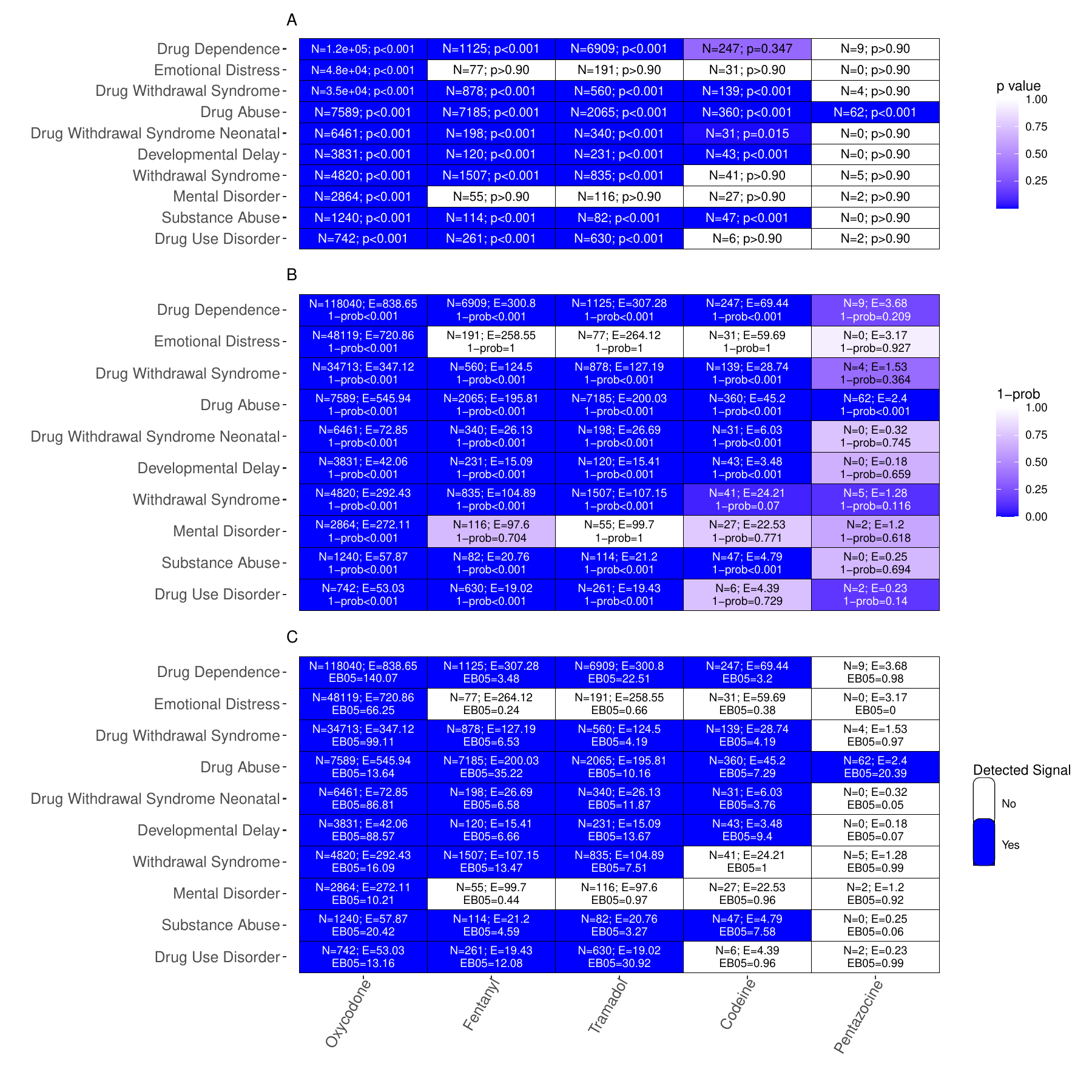}
    \caption{Signal detection results on the FAERS-opioid–mental data obtained from pseudo-LRT (panel A), general-gamma (panel B) and GPS (panel C). For each panel, rows correspond to the 10 prominent AEs, and columns correspond to the five opioids. Each cell displays observed report count $N$, expected null baseline $E$ (B, C) and respective measures of significance for corresponding AE-drug pair.} 
    \label{fig:faers-heatmap-comb}
\end{figure}

\begin{figure}[ht]
    \centering
    \includegraphics[width=\textwidth]{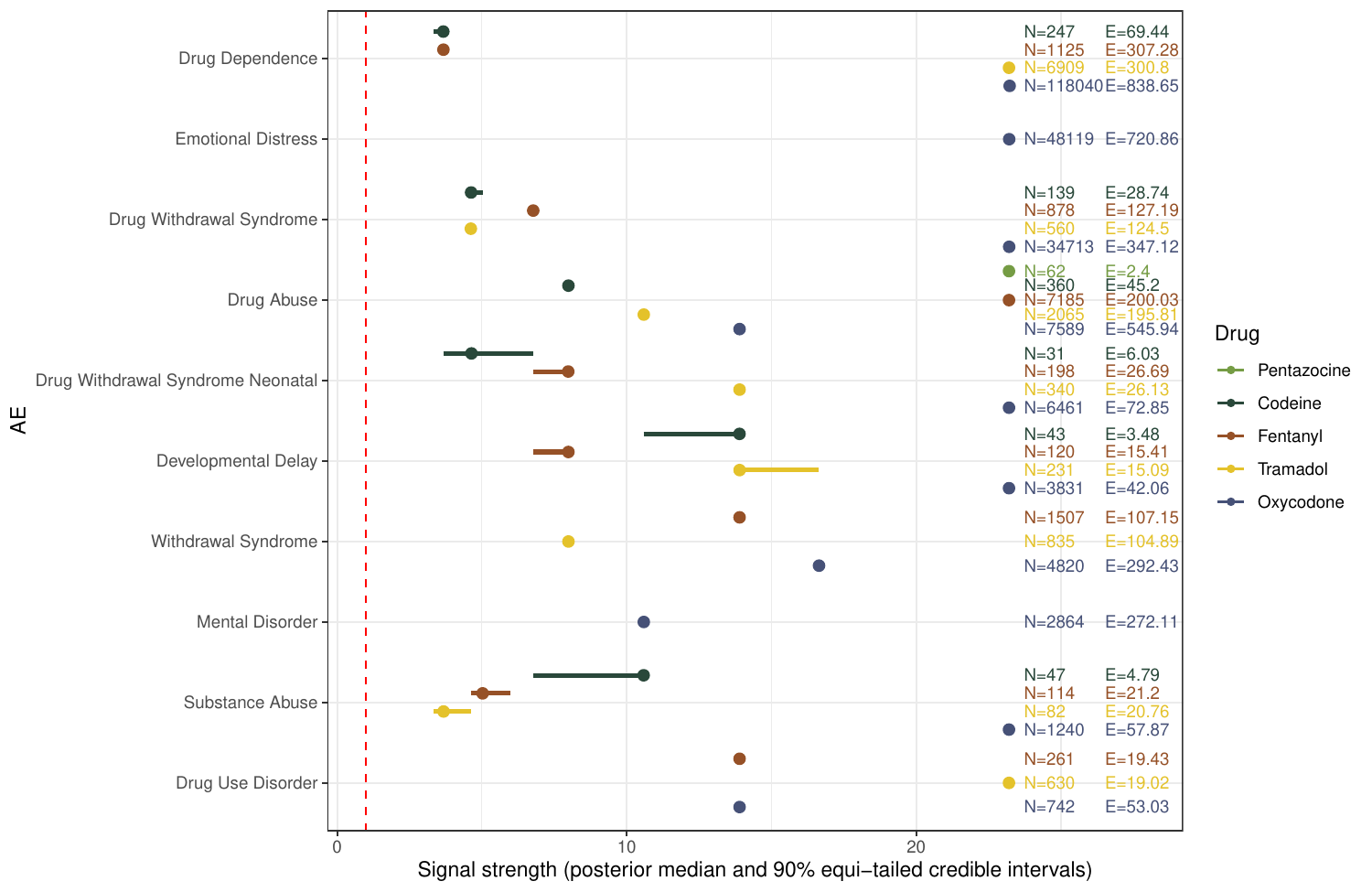}
    \caption{Nonparametric empirical Bayes (general-gamma) signal strength estimation results for the FAERS-opioid–mental data that visualizes empirical posterior inferences on top 10 prominent AEs across 5 opioid drugs (color-coded). Points and bars denote posterior medians and $90\%$ equi-tailed credible intervals for signal strength parameter $\{\lambda_{ij}\}$ corresponding to AE-drug pairs. The vertical red dotted line indicates the reference value of ``1''. The text on the right reports the observed counts and the expected counts under independence for each AE-drug pair.} 
    \label{fig:gg-faers-eyeplot}
\end{figure}

We conducted downstream SRS data mining analyses for two opioid data sets described in Section 3.1, VigiBase-opioid-mental and FAERS-opioid-mental. We consider the general-gamma, pseudo-LRT, and GPS methods, providing a comparison across frequentist and empirical Bayes methods. All computations were performed using R version 4.5.0 and the packages \textbf{pvLRT} (version 0.5.1), \textbf{pvEBayes} (version 0.2.2), and \textbf{openEBGM} (version 0.9.1). The results are illustrated with heatmaps and eyeplots by top AEs across opioid drugs. We also provide the most significant 10 AEs for each opioid drug in supplemental Tables S4-S9. 

Figure~\ref{fig:faers-heatmap-comb} presents heatmaps of the signal detection results for the FAERS-opioid-mental dataset, comparing the pseudo-LRT (panel A), general-gamma (panel B), and GPS (panel C) methods. The visualizations evaluate five opioid drugs against the top 10 most prominent AEs, which are ranked by their highest pseudo log-likelihood ratio statistic. To facilitate a direct comparison, the heatmaps for general-gamma and GPS results (panels B and C) focus on the same set of AEs. 

In Figure~\ref{fig:faers-heatmap-comb}A (pseudo-LRT), each cell reports the observed report count $N$, and the associated p-value for testing AE-drug association against the null hypothesis of independence is presented. An AE-drug pair with a p-value less than 0.05 is considered a detected signal. Oxycodone exhibits consistently significant association across the top AEs, with $p<0.001$. For Fentanyl, Tramadol, and Codeine, significant associations were discovered in most of the top AEs. In contrast, Pentazocine presents a different association pattern in this heatmap compared with other opioid drugs. It is non-significant to most of the top AEs presented in the heatmap except for ``Drug Abuse''. Pentazocine only has 4 AEs found with significant association by the pseudo-LRT method in FAERS-opioid–mental data, see supplemental Tables S4.  The Mayo Clinic \cite{MayoClinicPentazocineInjection2025}\cite{MayoClinicPentazocineNaloxoneOral2025} notes that Pentazocine is indicated for moderate to severe pain and may be used before surgery. Its oral product is co-formulated with naloxone, which blocks opioid effects, the “high” feeling. These can lead to fewer misuse-driven exposures. 

Figure~\ref{fig:faers-heatmap-comb}B presents the general-gamma results, including the observed report count $N$, expected null baseline $E$, and a significance measure denoted as "1-prob". This measure represents 1 minus posterior probability of being a signal,  $1 - p(\lambda > 1.001 \mid N)$ (see equation \eqref{eqn:signal-prob}), which is analogous to p-value in pseudo-LRT method; smaller values correspond to a stronger AE-drug association. As a default choice, we consider an AE-drug pair with "1-prob" smaller than 0.05 as a detected signal. Panel C shows the GPS results displaying the observed report count $N$, expected null baseline $E$, and significance measure "EB05", defined as the 5\% posterior quantile of the signal strength for a given AE-drug pair. Because the \textbf{openEBGM} documentation does not clearly specify a threshold for signal detection, we applied the decision rule recommended by the user guide of Oracle Life Sciences Empirica \cite{oracle_empirica_2025}, a commercial implementation of the GPS/MGPS method. Specifically, an AE-drug pair is classified as a detected signal if its "EB05" exceeds 2. In the heatmap, these significant signals are highlighted in blue, while non-significant pairs remain white. The signal detection results of general-gamma (panel B) and GPS (panel C) agree with the pseudo-LRT, showing a similar AE-drug association pattern for these opioid drugs on the FAERS-opioid-mental dataset.

Figure~\ref{fig:gg-faers-eyeplot} provides the signal strength estimation result of the general-gamma method on FAERS-opioid-mental data, which visualizes the empirical posterior inferences on the 10 prominent AEs (same set of AEs in Figure~\ref{fig:faers-heatmap-comb}) across 5 opioid drugs. The AE-drugs pairs determined as non-signals (posterior 5\% percentile of $\lambda$ smaller than 1.001) are excluded from this figure. The red dashed line denotes the null baseline of $\lambda = 1$, representing no association. For each AE-drug pair, the plotted dot and its associated horizontal line represent the posterior median and the equal-tailed 90\% posterior credible interval, respectively, for the signal strength parameter $\lambda$. The estimated signal strengths vary considerably across the AEs and drugs, reflecting a nuanced uncertain-aware quantification for signal strength that binary signal detection methods can not provide. Furthermore, the width of these intervals reflects the uncertainty associated with signal strength estimation; drug-AE pairs with smaller observed counts exhibit wider credible intervals. For example,  the AE-drug pairs ``Withdrawal Syndrome - Tramadol'' and ``Developmental Delay - Fentanyl'' have similar posterior median for signal strength $\lambda$. Since the latter AE-drug pair has a much smaller number of observations ($N=120$ vs. $N=835$), it shows a wider 90\% posterior credible interval. 

In addition, we note that the empirical Bayes construction of the general-gamma method induces a shrinkage effect that regularizes the posterior $\lambda_{ij}$ estimates, moderating extreme values. Because $O/E$ ratios are highly sensitive to relatively small expected baseline counts $E$, they are prone to severe inflation. By borrowing information from the entire dataset through an estimated prior distribution, the general-gamma method moderates noisy observations toward more conservative posterior estimates. For example, the AE-drug pair ``Developmental Delay - Oxycodone'' has a raw \(O/E = 3831/42.06 \approx 91.08\) and its posterior median is shrunk to less than 25 under the general-gamma model. In pharmacovigilance practice, this regularization prevents pharmacovigilance investigations from being misled by inflated signal strength estimations. 

%The AE-drug pairs that appeared in the eyeplot (Figure~\ref{fig:gg-faers-eyeplot}) are the same as those that appear in the heatmap (Figure~\ref{fig:gg-faers-heatmap}). Although they are all detected as signals, the estimated signal strength $\lambda$ implies that different levels of AE-drug association can lead to different medical/clinical implications.

%Figures~\ref{fig:vigi-heatmap-comb} and \ref{fig:gg-vigi-eyeplot} provide the signal detection (with pseudo-LRT, general-gamma and GPS) and signal strength estimation (general-gamma) results on the VigiBase-opioid-mental dataset. Overall, the VigiBase-opioid-mental signal detection results are similar to FAERS. The major is in ``Codeine'' drug column, where the general-gamma method consider all 10 AEs as detected signal while pseudo-LRT and GPS detect the same 6 AEs out of 10 AEs. We note that the decision rule used for general-gamma is equivalent to 5\% posterior quantile greater than 1.001 compared with 5\% posterior quantile greater than 2 that was used in the GPS method. The significant cutoff in the pseudo-LRT method is adaptively determined from SRS data by controlling type-I error under a specific level, while general-gamma and GPS signal detection thresholds are used at a fixed value. 

\begin{figure}[ht]
    \centering
    \includegraphics[width=\textwidth]{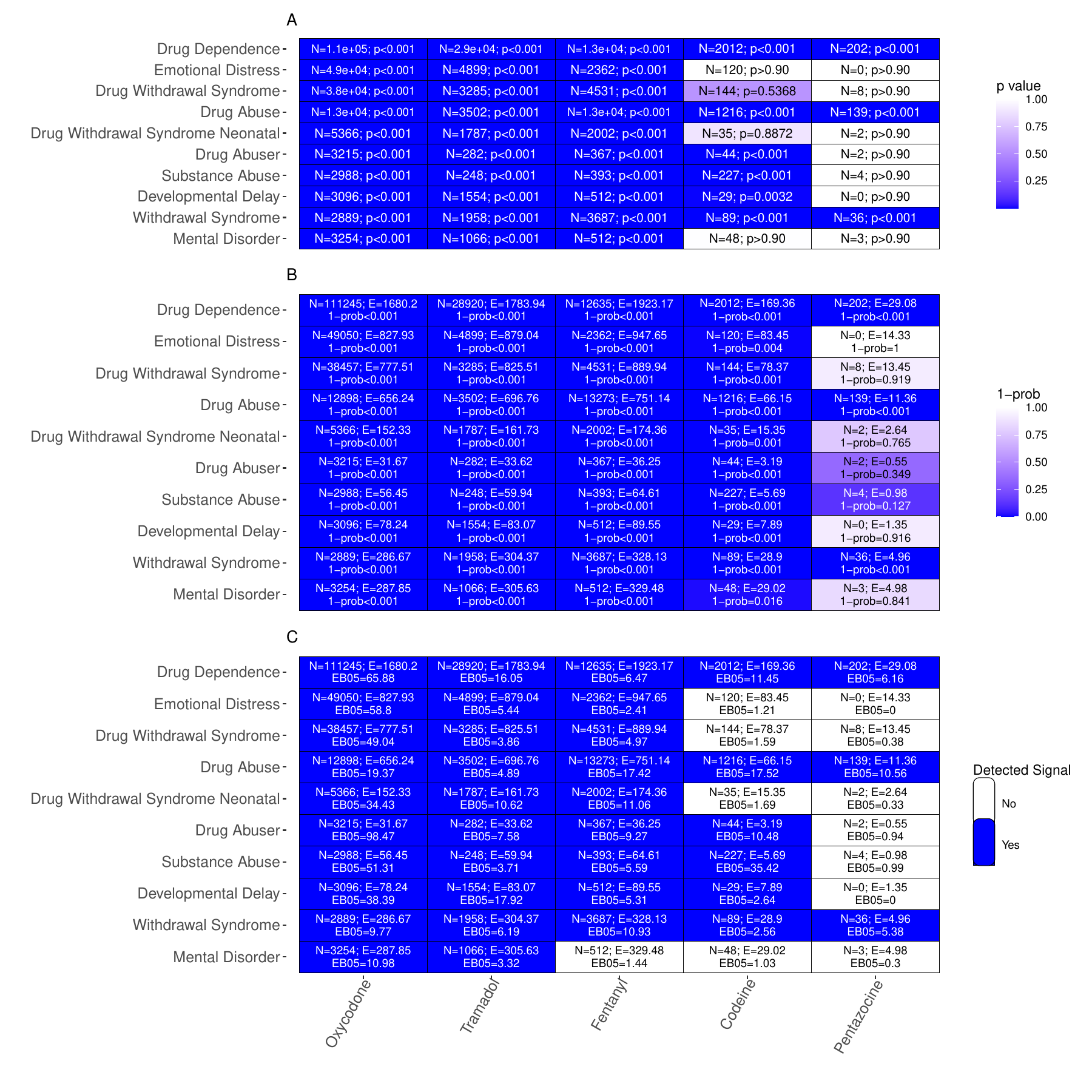}
    \caption{Signal detection results on the VigiBase-opioid–mental data obtained from pseudo-LRT (panel A), general-gamma (panel B) and GPS (panel C). For each panel, rows correspond to the 10 prominent AEs, and columns correspond to the five opioids. Each cell displays observed report count $N$, expected null baseline $E$ (B, C) and respective measures of significance for corresponding AE-drug pair.} 
    \label{fig:vigi-heatmap-comb}
\end{figure}

Figure~\ref{fig:vigi-heatmap-comb} presents the signal detection results (comparing pseudo-LRT, general-gamma, and GPS) for the VigiBase-opioid-mental dataset. Overall, the signal detection patterns in VigiBase-opioid-mental are highly similar to those observed in the FAERS-opioid-mental dataset. The major discrepancy occurs in the Codeine column, where the general-gamma method detects all 10 AEs as signals, whereas the pseudo-LRT and GPS methods detect the exact same 6 AEs. We note that the decision rule used for the general-gamma method ($1-\text{prob} < 0.05$) is mathematically equivalent to requiring the 5\% posterior quantile to be greater than 1.001. This is a much more permissive threshold compared to the GPS method that requires the 5\% posterior quantile (EB05) to be greater than 2. Furthermore, while the general-gamma and GPS methods rely on fixed, predefined thresholds, the significance cutoff in the pseudo-LRT method is adaptively determined from the SRS data to strictly control the Type-I error rate. This comparison highlights the impact of significance thresholds on signal detection outcomes. Existing literature discusses determining adaptive cutoffs by controlling the false discovery rate (FDR) \cite{muller2006fdr}, and this approach has been previously applied to the BCPNN and GPS methods \cite{ahmed2009bayesian}. Although PhViD\cite{ahmed_phvid_2016} implemented these Bayesian FDR controls for GPS, it has been archived by CRAN since May 2024 after failing automated compatibility checks, reflecting a lack of active maintenance since its last update in 2016. For contemporary Bayesian SRS data mining approaches, such as HDP and nonparametric empirical Bayes methods, this has not yet been rigorously studied.

Another noticeable difference is that the AE-drug pair ``Drug Dependence-Pentazocine'' is detected as a signal by all three methods in VigiBase data, while it is not a detected signal in FAERS data. Differences in SRS data mining results across datasets (FAERS vs. VigiBase) are expected even when the same drugs and AE families are selected. One possible reason is the differences in data access and preprocessing. Specifically, FAERS provides raw quarterly files, enabling investigators to focus on a precise study window (e.g., 2014Q3–2024Q4). In contrast, publicly accessible VigiBase data does not support such temporal filtering. Consequently, long-marketed drugs may reflect a cumulative exposure history in VigiBase that exceeds the specific window applied to FAERS. This usually happens in the reference drug column, which consists of common daily-use drugs (e.g., aspirin). Furthermore, disparities in underlying populations and medical regulation also contribute to heterogeneity in reporting patterns. FAERS reflects the U.S. population, whereas VigiBase aggregates global reports, encompassing diverse healthcare systems, opioid availability, prescribing norms, and reporting obligations. While these differences inherently introduce batch effects across databases, a detailed investigation of their impact is beyond the scope of this article.

\begin{figure}[ht]
    \centering
    \includegraphics[width=\textwidth]{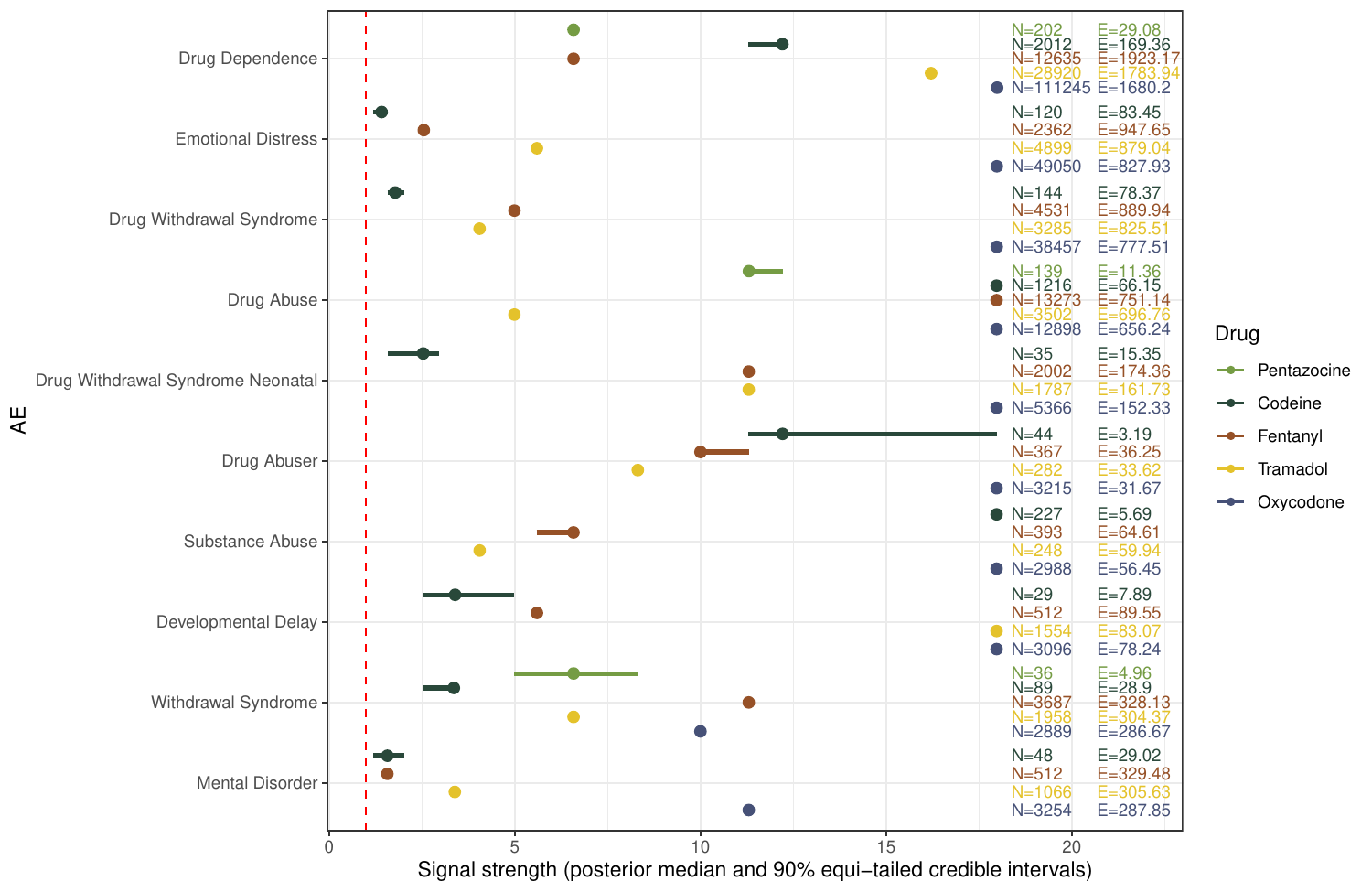}
    \caption{Nonparametric empirical Bayes (general-gamma) signal strength estimation results for the vigiBase-opioid–mental data that visualizes empirical posterior inferences on top 10 prominent AEs across 5 opioid drugs (color-coded). Points and bars denote posterior medians and $90\%$ equi-tailed credible intervals for signal strength parameter $\{\lambda_{ij}\}$ corresponding to AE-drug pairs. The vertical red dotted line indicates the reference value of ``1''. The text on the right reports the observed counts and the expected counts under independence for each AE-drug pair.} 
    \label{fig:gg-vigi-eyeplot}
\end{figure}

Figure~\ref{fig:gg-vigi-eyeplot} displays signal strength estimations (general-gamma method) for the VigiBase opioid-mental dataset. The plot displays the posterior median (dots) and 90\% equal-tailed credible intervals (lines) for the top 10 AEs across five opioid drugs. The red dashed line represents the null baseline ($\lambda = 1$). Similar to the signal strength estimation results in the FAERS-opioid-mental dataset, we observe varying estimated signal strengths across AEs and drugs, with their associated uncertainty reflected by the posterior credible intervals.

% \begin{figure}[ht]
%     \centering
%     \includegraphics[width=\textwidth]{fig/lrt_vigi_heatmap.pdf}
%     \caption{Pseudo-LRT signal detection results for the Vigi-opioid–mental data. Rows correspond to top 10 AEs and columns correspond to the five opioids. AE rows are ordered by the largest pseudo log-likelihood ratio statistic (logLR) in each row. Each cell displays the observed report count n, logLR, and the associated p-value for testing AE-drug association against the null hypothesis of independence; the cell filling color represents the p-value (deeper colors indicate stronger evidence of association).} 
%     \label{fig:lrt-vigi}
% \end{figure}

% \begin{figure}[ht]
%     \centering
%     \includegraphics[width=\textwidth]{fig/gg_vigi_heatmap.pdf}
%     \caption{Nonparametric empirical Bayes (general-gamma) signal detection results for the vigi-opioid–mental data. Rows correspond to top 10 AEs and columns correspond to the five opioids. AE rows are ordered by the largest posterior probability $p(\lambda > 1.001 \mid N)$ in each row. Each cell displays the observed report count $N$, expected null baseline $E$, and the associated posterior probability of being a signal for the corresponding AE-drug association; the cell filling color represents the posterior probability of being a signal (deeper colors indicate stronger evidence of association).} 
%     \label{fig:gg-vigi-heatmap}
% \end{figure}

\section{Conclusion}

In this paper, we reviewed contemporary SRS data mining methods in pharmacovigilance. We proposed a procedure to obtain an SRS frequency table for a database that is only publicly available to aggregated AE-drug count, e.g., VigiBase and EudraVigilance. Based on this, we obtained four SRS datasets (VigiBase-opioid, VigiBase-opioid-mental, FAERS-opioid, and FAERS-opioid-mental). We illustrated the use of two representative methods (pseudo-LRT and general-gamma) through VigiBase-opioid-mental and FAERS-opioid-mental datasets. 

Across both databases, the three methods largely agreed on the detected AE-drug pairs, while providing different summaries. The pseudo-LRT offered a signal/non-signal dichotomy (signal detection) for AE-drug pairs under a rigorous frequentist hypothesis testing framework, while GPS provides empirical Bayes signal detection results. Notably, the nonparametric empirical Bayes method (general-gamma) not only detects signal based on posterior probabilities, but also provide interpretable posterior estimates of signal strength ($\lambda$), yielding nuanced uncertainty-aware quantification for signal strengths. These are particularly valuable in a medical or clinical context. We note that, as with all spontaneous reporting analyses, AE-drug associations reflect disproportional reporting rather than causal effects and may be influenced by under-reporting, lack of proper controls, inaccuracies in measuring drug use, the presence of selection bias, and confounding. Differences between FAERS and VigiBase results shown in Section \ref{sec:disproportionality-analysis} are expected. They can arise from (i) database-specific data access and preprocessing (e.g., time window alignment, aggregation constraints); (ii) heterogeneity in underlying populations and prescribing patterns, and (iii) variation in medical regulation and reporting practices across different geographic regions. 

Finally, we highlight several challenges in SRS data mining that need further investigation: (1) the specific impact that reference row and column during data construction has on downstream analysis; (2) the need for statistical methods to handle temporal alignment and the data heterogeneity discussed in Section \ref{sec:opioid-datasets}; and (3) the lack of standardized, adaptive cutoff determination for contemporary Bayesian signal detection frameworks. 

\section*{Acknowledgment}

The research was funded by a KALEIDA Health Foundation award (award number 82114) to Dr. Markatou that supported the work of the first author. 

\clearpage

\bibliographystyle{unsrt}  
%\bibliography{references}  %%% Remove comment to use the external .bib file (using bibtex).
%%% and comment out the ``thebibliography'' section.

%%% Comment out this section when you \bibliography{references} is enabled.
\bibliography{references}

\end{document}